\documentclass[aps,showpacs,11pt]{revtex4}
\usepackage{dcolumn}
\usepackage{graphicx}
\usepackage{amsmath}
\usepackage{amsfonts}
\usepackage{amssymb}
\usepackage{psfrag}
\usepackage{wrapfig}
\usepackage{subfigure}
\usepackage{makeidx}
\usepackage{bm}
\usepackage{epsf}
\usepackage{cases}
\usepackage{hyperref}
\usepackage{multirow}
\makeatletter

\newcommand{\Rmnum}[1]{\expandafter\@slowromancap\romannumeral #1@}
\makeatother

\begin{document}

\title{Reentrant phase transitions of higher-dimensional AdS black holes in dRGT massive gravity
}


\author{De-Cheng Zou}
\email{dczou@yzu.edu.cn}
\author{Ruihong Yue}
\email{rhyue@yzu.edu.cn}
\affiliation{Center for Gravitation and Cosmology,
College of Physical Science and Technology, Yangzhou University, Yangzhou 225009, China}

\author{Ming Zhang}
\email{shakellar@126.com}
\affiliation{Faculty of Science, Xi'an Aeronautical University, Xi'an 710077,China}

\begin{abstract}

We study the $P-V$ criticality and phase transition in the extended phase space of
anti-de Sitter (AdS) black holes in higher-dimensional
de Rham, Gabadadze and Tolley (dRGT) massive gravity,
treating the cosmological constant as pressure and the corresponding
conjugate quantity is interpreted as thermodynamic volume.
Besides the usual small/large black hole phase transitions,
the interesting thermodynamic phenomena of reentrant phase transitions (RPTs)
are observed for black holes in all $d\geq6$-dimensional spacetime
when the coupling coefficients
$c_i m^2$ of massive potential satisfy some certain conditions.
\end{abstract}

 \pacs{04.50.Kd, 04.70.Dy, 04.50.Gh}
 
\keywords{dRGT massive gravity,  $P-V$ criticality , higher dimensional black holes}


\maketitle

\section{Introduction}
\label{intro}

Einstein's general relativity (GR) is a relativistic theory of gravity where the
graviton is a massless spin-2 particle \cite{Gupta:1954zz,Weinberg:1965rz,Feynman:1996kb}.
It is also the current description of gravitation in modern physics and has significant
astrophysical implications. Nevertheless, whether there exist a consistent extension of GR by
a mass term is a basic challenge of classical field theory, since the open questions such
as the old cosmological constant problem and the origin of the late-time acceleration
of the Universe remain behind the puzzles at the interface between gravity/cosmology
and particle physics.
In general, by adding generic mass terms for the gravitons on the given background
usually brings about various instabilities for the gravitational theories, sometimes on the
nonlinear level. A new nonlinear massive gravity theory was proposed by
de Rham, Gabadadze and Tolley (dRGT) \cite{deRham:2010ik,deRham:2010kj,Hinterbichler:2011tt},
where the Boulware-Deser ghost \cite{Boulware:1973my} was eliminated by introducing
higher order interaction terms in the action.
Until now, a nontrivial black hole solution with a Ricci flat horizon has been
constructed by Vegh \cite{Vegh:2013sk,Adams:2014vza} in four-dimensional
dRGT massive gravity. Later, the spherically symmetric solutions were also
addressed in \cite{Nieuwenhuizen:2011sq,Brito:2013xaa,Do:2016abo}, and
the corresponding charged
black hole solution was found in \cite{Berezhiani:2011mt}, including its
bi-gravity extension \cite{Babichev:2014fka,Do:2016uef}.
Moreover, the charged AdS black hole solution in higher-dimensional dRGT
massive gravity, and its corresponding thermodynamics and phase structure
in the grand canonical and canonical ensembles were also presented in \cite{Cai:2014znn}.
Ge et al. \cite{Ge:2014aza} examined the relations between dynamical
instabilities and thermodynamic instabilities in the dRGT massive gravity.

Recently, the study of thermodynamics in AdS black
holes has been generalized to the extended phase
space, where the cosmological constant is regarded
as a variable and also identified with thermodynamic pressure
\cite{Dolan:2011xt,Dolan:2010ha}
\begin{eqnarray}
P=-\frac{\Lambda}{8\pi}=\frac{(d-1)(d-2)}{16\pi l^2}\label{eq:1a}
\end{eqnarray}
in the geometric units $G_N=\hbar=c=k=1$. Here $d$ stands for the number of spacetime
dimensions and $l$ denotes the AdS radius. In this case, the variation
of the cosmological constant is included in the first law
of black hole thermodynamics, which ensures the consistency between
the first law of black hole thermodynamics and the Smarr formula.
In \cite{Caceres:2015vsa}, it was pointed out that the extended phase space
can be interpreted as an RG-flow in the space of field theories,
where isotherm curves codify how the number of degrees of freedom
$N$ (or the central charge $c$) runs with the energy scale.
Moreover, the variation of cosmological constant could be
corresponded to variation of number of the colors in Yang-Mills
theory residing on the boundary spacetime \cite{Creighton:1995au,Gibbons:1996af}.
In the extended phase space, the charged AdS black hole admits a more direct
and precise coincidence between the first order small/large black holes
(SBH/LBH) phase transition and Van der Waals liquid-gas phase transition,
and both systems share the same critical exponents near the
critical point \cite{Kubiznak:2012wp}. As a result, the analogy between the
charged AdS black hole and the Van der Waals system becomes more complete.
More discussions in various gravity theories can be found in
\cite{Hansen:2016ayo,Dutta:2013dca,
Xu:2013zea,Dehghani:2014caa,Hendi:2012um,Xu:2014kwa,Zou:2013owa,Cai:2013qga,
Xu:2014tja,Zhao:2014raa,Altamirano:2014tva,Mo:2014qsa,Zou:2014mha,
Zhang:2015ova,Zhang:2014eap,Zhang:2014uoa,Majhi:2016txt,Lee:2014tma,
Hendi:2016usw,Kubiznak:2016qmn,Hendi:2016njy,Kuang:2016caz,Miao:2016ipk,Cadoni:2017ktd}.
In this direction, some investigations for thermodynamics of
AdS black holes in the dRGT massive gravity have been generalized to the extended phase
space \cite{Hendi:2015hoa,Xu:2015rfa,Zhang:2016fxj,Ghosh:2015cva,Prasia:2016fcc,Hendi:2017fxp},
which revealed the existence of Van der Waals-like SBH/LBH phase transition.
In addition, a link between the Van der Waals-like SBH/LBH
phase transition and quasinormal modes (QNMs)
has established in four \cite{Liu:2014gvf} and higher \cite{Chabab:2016cem} dimensional
Reissner-Nordstr$\ddot{o}$m AdS black hole, including time-domain
profiles \cite{Chabab:2017knz}, and higher-dimensional charged black hole in
the presence of Weyl coupling \cite{Mahapatra:2016dae}.
In terms of AdS/CFT, holographic entanglement entropy (HEE),
Wilson loop, and two point correlation function also provide useful tools
to probe the Van der Waals-like SBH/LBH phase transition \cite{Dey:2015ytd,Li:2017gyc,Zeng:2015wtt,Zeng:2016aly,Sun:2016til,Zeng:2016sei}.

Recently, Ref.~\cite{Gunasekaran:2012dq} firstly recovered the existence of
intermediate/small/large phase transitions in the four- dimensional Born-Infeld-AdS
black hole, which is reminiscent of reentrant phase transitions (RPTs)
observed for multicomponent fluid systems, ferroelectrics, gels, liquid crystals,
and binary gases, e.g.,\cite{Narayanan}. A system undergoes an RPT if a monotonic
variation of any thermodynamic quantity results in two
(or more) phase transitions such that the final state is
macroscopically similar to the initial state. Moreover, this
RPT also appears in the higher-dimensional rotating AdS black holes \cite{Altamirano:2013ane,Kubiznak:2015bya}, five-dimensional hairy AdS black
hole \cite{Hennigar:2015wxa}, and higher-dimensional Gauss-Bonnet AdS black hole \cite{Frassino:2014pha,Hennigar:2016ekz}. It is interesting to generalize
the discussion to the black holes in the dRGT massive gravity. In this paper,
we will report the finding of interesting RPTs
in all $d\geq6$-dime-nsional black holes when the coupling coefficients
$c_i m^2$ of massive potential satisfy some certain conditions.

This paper is organized as follows. In Sect.~\ref{2s}, we review the
thermodynamics of massive gravity black holes in the extended phase space.
In Sect.~\ref{3s}, we study the critical behavior of higher-dimensional
AdS black hole in context of $P-V$ criticality and phase diagrams.
We end the paper with closing remarks in Sect.~\ref{4s}.

\section{Thermodynamics of higher-dimensional AdS black hole in dRGT massive gravity}
\label{2s}

We start with the action of higher-dimensional dRGT massive gravity in presence
of a negative cosmological constant
\begin{eqnarray}
\mathcal{I}=\frac{1}{16\pi}\int{d^{d} x\sqrt{-g}\left[R-2\Lambda+m^2\sum_{i=1}^{4}c_{i}{\cal U}_{i}(g,f)\right]},\label{action}
\end{eqnarray}
where the last four terms are the massive potential associate with graviton mass $m$,
$c_i$ are constants and $f$ is a fixed rank-2 symmetric tensor.
Moreover, ${\cal U}_{i}$ are symmetric polynomials of the eigenvalues
of the $d\times d$ matrix ${\cal K}^\mu_{\nu}\equiv\sqrt{g^{\mu\alpha}f_{\alpha\nu}}$
\begin{eqnarray}
{\cal U}_{1}&=&[\cal K],\nonumber\\
{\cal U}_{2}&=&[{\cal K}]^2-[{\cal K}^2],\nonumber\\
{\cal U}_{3}&=&[{\cal K}]^3-3[{\cal K}][{\cal K}^2]+2[{\cal K}^3],\nonumber\\
{\cal U}_{4}&=&[{\cal K}]^4-6[{\cal K}^2][{\cal K}]^2
+8[{\cal K}^3][{\cal K}]\nonumber\\
&&+3[{\cal K}^2]^2-6[{\cal K}^4].\label{ac}
\end{eqnarray}
The square root in ${\cal K}$ is understood as the matrix square root, ie.,
$(\sqrt{A})^{\mu}_{~\nu}(\sqrt{A})^{~\nu}_{\lambda}=A^\mu_{~\lambda}$, and the
rectangular brackets denote traces $[{\cal K}]={\cal K}^{\mu}_{~\mu}$.

Consider the metric of $d$-dimensional spacetime in the following form
\begin{eqnarray}
ds^2=-f(r)dt^2+\frac{1}{f(r)}dr^2+r^2h_{ij}dx^idx^j,\label{metric}
\end{eqnarray}
where $h_{ij}dx^idx^j$ is the line element for
an Einstein space with constant curvature $(d-2)(d-3)k$.
The constant $k$ characterizes the geometric property of
black hole horizon hypersurface, which takes values $k=0$ for flat, $k=-1$ for
negative curvature and $k=1$ for positive curvature, respectively.

By using the reference metric
\begin{eqnarray}
f_{\mu\nu}=diag(0,0,c_0^2h_{ij}),
\end{eqnarray}
the metric function $f(r)$ is obtained as \cite{Cai:2014znn}
\begin{eqnarray}
f(r)&&=k+c_0^2c_2m^2+\frac{16\pi P}{(d-1)(d-2)}r^2+\frac{c_0c_1m^2}{d-2}r\nonumber\\
&&-\frac{16\pi M}{(d-2)V_{d-2} r^{d-3}}+\frac{(d-3)c_0^3c_3m^2}{r}\nonumber\\
&&+\frac{(d-3)(d-4)c_0^4c_4m^2}{r^2}.\label{solution}
\end{eqnarray}
Here $c_0$ is a positive constant, $V_{d-2}$ is the volume of space spanned by
coordinates $x^i$, and $M$ is the black hole mass. It is necessary to point out that
the terms $c_3m^2$ and $c_4m^2$ only appear
in the black hole solutions for $d\geq5$ and $d\geq6$, respectively \cite{Cai:2014znn}.
When $m\rightarrow0$, namely, without the massive potential, Eq.~(\ref{solution})
reduces to the $d$-dimensional Schwarzschild AdS (SAdS) black hole solution.

In terms of the radius of the horizon $r_+$, the mass $M$, Hawking temperature $T$ and
entropy $S$ of black holes can be written as
\begin{eqnarray}
M&=&\frac{(d-2)V_{d-2}r_+^{d-3}}{16\pi}\left[k+\frac{16\pi P}{(d-1)(d-2)}r_+^2+\frac{c_0c_1m^2r_+}{d-2}\right.\nonumber\\
&&\left.+c_0c_2m^2+\frac{(d-3)c_0^3c_3m^2}{r_+}
+\frac{(d-3)(d-4)c_0^4c_4m^2}{r_+^2}\right],\nonumber\\
T&=&\frac{f'(r_+)}{4\pi}
=\frac{1}{4\pi r_+}\left[(d-3)k+\frac{16\pi P}{d-2}r_+^2+c_0c_1m^2r_{+}\right.\nonumber\\
&&\left.+(d-3)c_0^2c_2m^2+\frac{(d-3)(d-4)c_0^3c_3m^2}{r_+}\right.\nonumber\\
&&\left.+\frac{(d-3)(d-4)(d-5)c_0^4c_4m^2}{r_+^2}\right],\nonumber\\
S&=&\frac{V_{d-2}}{4} r_+^{d-2}. \label{eq:5a}
\end{eqnarray}
The black hole mass $M$ can be considered as the
enthalpy rather than the internal energy of the gravitational system.
Moreover, the first law of black hole thermodynamics and Smarr
relation are given by \cite{Xu:2015rfa}
\begin{eqnarray}
dM&=&TdS+VdP+\frac{c_0m^2V_{d-2}r_+^{d-2}}{16\pi}dc_1\nonumber\\
&&+\frac{(d-2)c_0^2m^2V_{d-2}r_+^{d-3}}{16\pi}dc_2\nonumber\\
&&+\frac{(d-2)(d-3)c_0^3m^2V_{d-2}r_+^{d-4}}{16\pi}dc_3\nonumber\\
&&+\frac{(d-2)(d-3)(d-4)c_0^4m^2V_{d-2}r_+^{d-5}}{16\pi}dc_4,\label{flbh}\\
(d-3)M&=&(d-2)TS-2VP-\frac{c_0c_1m^2V_{d-2}}{16\pi}r_+^{d-2}\nonumber\\
&&+\frac{(d-2)(d-3)c_0^3c_3m^2V_{d-2}}{16\pi}r_+^{d-4}\nonumber\\
&&+\frac{(d-2)(d-3)(d-4)c_0^4c_4m^2V_{d-2}r_+^{d-5}}{8\pi},\label{srbh}
\end{eqnarray}
where $V_{d-2}$ denotes the thermodynamic volume and equals
to $\frac{V_{d-2}}{d-1}r_+^{d-1}$.

\section{Critical behaviors of higher-dimensional AdS black holes}
\label{3s}

\subsection{Equation of state}

For further convenience, we denote
\begin{eqnarray}
&&\hat{T}=T-\frac{c_0c_1m^2}{4\pi},\quad w_2=-\frac{k+c_0^2c_2m^2}{8\pi},\nonumber\\
&&w_3=-\frac{c_0^3c_3m^2}{8\pi},\quad w_4=-\frac{c_0^4c_4m^2}{8\pi},\label{wcoef}
\end{eqnarray}
Here $\hat{T}$ denotes the shifted temperature and can be negative according
to the value of $c_0c_1m^2$.
Then the equation of state of the black hole can be obtained from Eq.~(\ref{eq:5a})
\begin{eqnarray}
P&=&\frac{d-2}{4r_+}\left[\hat{T}+\frac{2(d-3)w_2}{r_+}+\frac{2(d-3)(d-4)w_3}{r_+^2}\right.\nonumber\\
&&\left.+\frac{2(d-3)(d-4)(d-5)w_4}{r_+^3}\right].\label{eos}
\end{eqnarray}
To compare with the Van der Waals fluid equation, we can translate
the ``geometric" equation of state to physical one by identifying the specific
volume $v$ of the fluid with the radius of the horizon of black hole as $v=\frac{4r_+}{d-2}$.
Evidently, the specific volume $v$ is proportional to the horizon radius $r_+$,
therefore we will just use the radius of the horizon in the equation of
state for the black hole hereafter in this paper.

We know that the critical point occurs when $P$
has an inflection point,
\begin{eqnarray}
\frac{\partial P}{\partial r_+}\Big|_{\hat{T}=\hat{T}_c, r_+=r_c}
=\frac{\partial^2 P}{\partial r_+^2}\Big|_{\hat{T}=\hat{T}_c, r_+=r_c}=0,
\label{eq:15a}
\end{eqnarray}
where the subscript stands for the quantities at the critical point.
The critical shifted temperature is obtained
\begin{eqnarray}
\hat{T}_{c}=-\frac{2(d-3)}{r_c}\left[2w_2+\frac{3(d-4)w_3}{r_c}
+\frac{4(d-4)(d-5)w_4}{r_c^2}\right],\nonumber\\\label{Teos}
\end{eqnarray}
and the equation for the critical horizon radius $r_c$ is given by
\begin{eqnarray}
6(d-4)(d-5)w_4+3(d-4)w_3r_c+w_2r_c^2=0.\label{eq:7a}
\end{eqnarray}
One can easily find that in four-dimensional spacetime $(d=4,w_3=w_4=0)$,
the absence of positive solution of Eq.~(\ref{eq:7a})
indicates that no criticality can occur \cite{Xu:2015rfa}.
A similar situation also occurs in
the $d$-dimensional Schwarzschild
AdS black hole ($m\rightarrow0$), since there does not exist any
real root of Eq.~(\ref{eq:7a}) with $w_3=w_4=0$.

We further discuss the critical behaviors of higher- dimensional
$(d\geq5)$ AdS black hole
when $w_2\neq0$ and $w_3\neq0$. When setting $w_4=0$, one have
\begin{eqnarray}
r_{c}&=&-\frac{3(d-4)w_3}{w_2},\quad
\hat{T}_{c}=\frac{2(d-3)w_2^2}{3(d-4)w_3},\nonumber\\
P_{c}&=&-\frac{(d-2)(d-3)w_2^3}{54(d-4)^2w_3^2}.\label{eq:8a}
\end{eqnarray}
Note that the critical behavior occurs only when $w_2<0$ and $w_3>0$.
We can easily find an interesting relation among the critical
pressure $P_c$, temperature $\hat{T}_c$ and horizon radius $r_c$
\begin{eqnarray}
\frac{P_{c}r_c}{\hat{T}_{c}}=\frac{d-2}{12}.\label{eq:9a}
\end{eqnarray}
For $d=5$, Eqs.~(\ref{eq:8a})(\ref{eq:9a})
reduce to the equations described in \cite{Xu:2015rfa}.

With regard to the case of $w_4\neq0$, which only appears for $d\geq6$,
the direct solution of Eq.~(\ref{eq:7a}) reads
\begin{eqnarray}
r_{c1,2}&=&\frac{1}{2w_2}\left[\pm\sqrt{9(d-4)^2w_3^2
-24(d-4)(d-5)w_2w_4}\right.\nonumber\\
&&\left.-3(d-4)w_3\right]\equiv\frac{\chi_{\pm}}{2w_2}\label{eq:18a}
\end{eqnarray}
if $3(d-4)w_3^2\geq8(d-5)w_2w_4$. In this case, $r_{c1}$ and $r_{c2}$
correspond to the $``-"$ and $``+"$ branches, respectively. The condition of $r_{c1,2}>0$
crucially depends on the dimension of spacetime and values of $w_2$, $w_3$ and $w_4$.

For $w_2<0$, the positivity of solution $r_{c1}$ leads to $w_3<0$ and $w_4>0$
or $w_3>0$. In order that $r_{c2}$ be positive, it requires
an additional constraint: $w_3>0$ and $w_4<0$.
By substituting the solutions $r_{c1,2}$ (\ref{eq:18a})
into Eqs.~(\ref{eos}) and (\ref{Teos}), we obtain
\begin{eqnarray}
&&\hat{T}_{c1,2}=\frac{8(d-3)(d-4)w_2^2\left[16(d-5)w_2w_4
+3w_3\chi_{\pm}\right]}{\chi_{\pm}^3},\label{Tw2}\\
&&P_{c1,2}=\frac{\pm4(d-2)(d-3)(d-4)w_2^3\left[6(d-5)w_2w_4
+w_3\chi_{\pm}\right]}{\chi_{\pm}^4}.\nonumber\\
\label{Pw2}
\end{eqnarray}
Due to the shifted temperature $\hat{T}_{c1,2}$ can be negative;
here we only evaluate the results of $P_{c1,2}$.
When $P_{c1}>0$, we shall keep $w_3<0$ and $w_4>0$ or $w_3>0$. On the other hand,
taking $w_3>0$, $w_4<0$ and $\frac{3(d-4)w_3^2}{8(d-5)w_4}<w_2<\frac{3(d-4)w_3^2}{9(d-5)w_4}$
lead to $P_{c2}>0$. As a result, two critical points
($r_{c1,2}>0$ and $P_{c1,2}>0$) will appear
in the range of $\frac{3(d-4)w_3^2}{8(d-5)w_4}<w_2<\frac{3(d-4)w_3^2}{9(d-5)w_4}$
with $w_3>0$ and $w_4<0$.

\begin{figure}[htb]
\centering
\subfigure[$w_4=0.1$]{ 
  \includegraphics{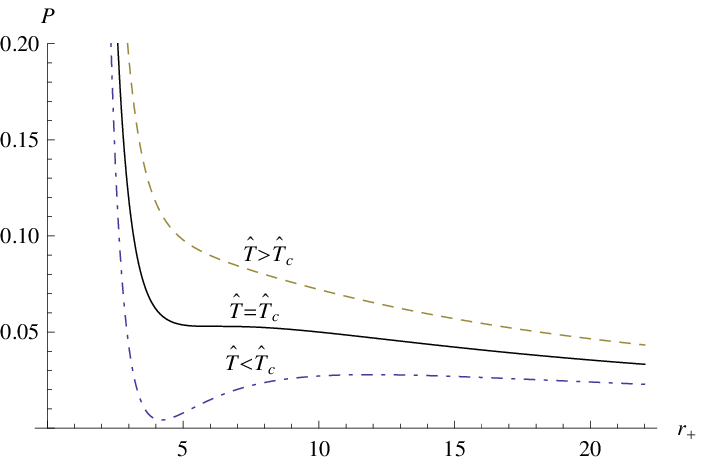}}%
\hfill%
\subfigure[$w_4=-0.7$]{ 
  \includegraphics{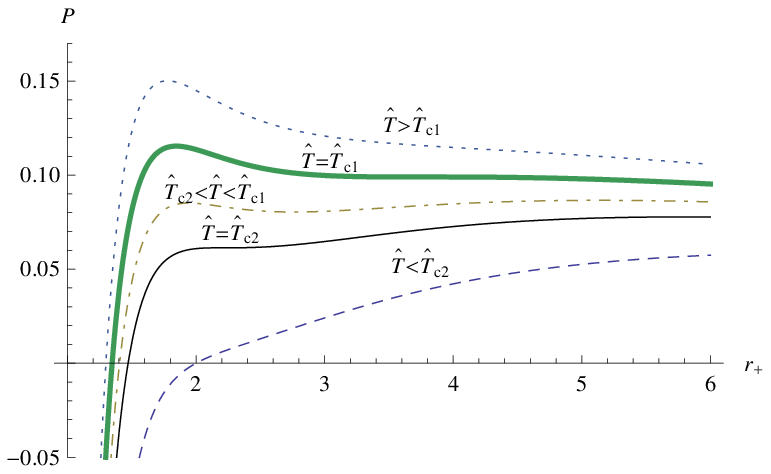}}%
  \caption{\textbf{ The $P-r_+$ diagrams of six-dimensional AdS black holes
for $w_2=-1$ and $w_3=1$.} (a) The upper dashed line corresponds to the idea gas phase
behaviour for $\hat{T}>\hat{T}_c$. The critical temperature case $\hat{T}=\hat{T}_c$ is denoted by the solid
line. The line below is with temperatures smaller than the critical temperature. We have
$\hat{T}_c=0.9788$ in $(a)$. In $(b)$, we have now two critical points at positive pressure. The
upper one has higher radius, temperature, and mass.
We have $\hat{T}_{c1}=1.27718$ and $\hat{T}_{c2}=1.1718$ in $(b)$. }\label{fig1}
\end{figure}

\begin{figure}[htb]
\centering
\subfigure[$w_2=-1$, $w_3=-1$ and $w_4=0.6$]{ 
\includegraphics{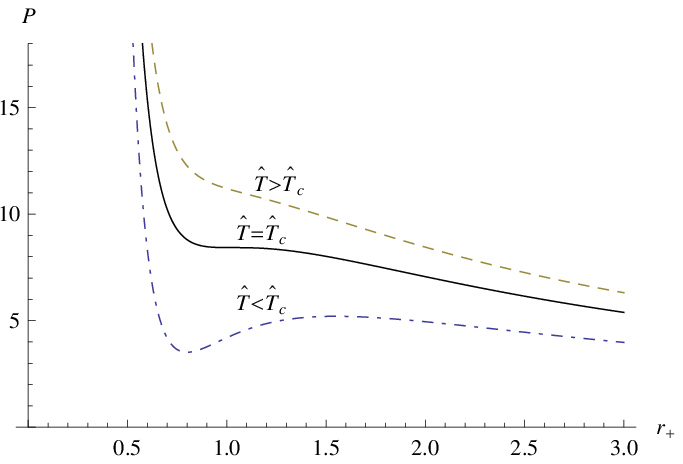}}%
\hfill%
\subfigure[$w_2=1$, $w_3=-1$ and $w_4=0.5$]{ 
\includegraphics{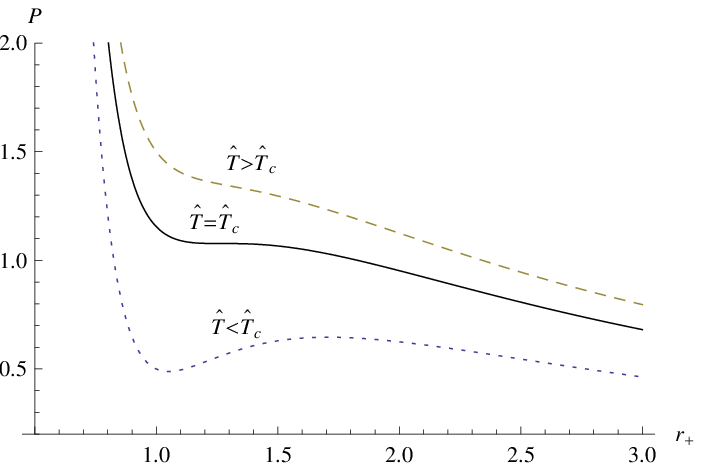}}%
\caption{ \textbf{The $P-r_+$ diagrams of six-dimensional AdS black holes.}
The upper dashed line corresponds to the idea gas phase behavior for $\hat{T}>\hat{T}_c$.
The critical temperature case $\hat{T}=\hat{T}_c$ is denoted by the solid line.
The lines below is with temperatures smaller than the critical temperature.
We have $\hat{T}_c=19.2290$ in $(a)$
and $\hat{T}_c=1.1547$ in $(b)$.}\label{fig2}
\end{figure}

In case of $w_2>0$, there is only one critical point ($r_{c1}>0$ and $P_{c1}>0$)
for $w_3<0$, $w_4>0$ and $w_2<\frac{(d-4)w_3^2}{3(d-5)w_4}$.
We summarize the critical points in the Table~\ref{sphericcase}.
The corresponding $P-r_+$ diagrams for $d=6$ are
displayed in Figs.{\ref{fig1}} and {\ref{fig2}}.

\begin{table}
\caption{The behaviour of critical points for different values of coupling
constants when $d\geq6$.}
\label{sphericcase}     
\begin{tabular}{llllll}
  \hline
\multicolumn{1}{|c|}{Parameters} & \multicolumn{3}{|c|}{$w_2<0$} & \multicolumn{1}{|c|}{$w_2>0$}\\ \hline
\multicolumn{1}{|c|}{$w_3$} & \multicolumn{2}{|c|}{$w_3>0$}&\multicolumn{1}{|c|}{$w_3<0$}&\multicolumn{1}{|c|}{$w_3<0$} \\ \hline
\multicolumn{1}{|c|}{$w_4$} &\multicolumn{1}{|c|}{$w_4\geq0$} & \multicolumn{1}{|c|}{$w_4<0$} &
\multicolumn{1}{|c|}{$w_4\geq0$} & \multicolumn{1}{|c|}{$w_4>0$} \\ \hline
\multicolumn{1}{|c|}{Number of critical
point}& \multicolumn{1}{|c|}{One} & \multicolumn{1}{|c|}{Two} &
\multicolumn{1}{|c|}{One} & \multicolumn{1}{|c|}{One}   \\ \hline
\end{tabular}
\end{table}

To study the possible phase transitions in the system,
let us now turn to the expression for the Gibbs free energy.

\subsection{Gibbs free energy}

The behavior of the free energy $G$ is
important to determine the thermodynamic phase
transition. The free energy $G$ obeys the thermodynamic relation
\begin{eqnarray}
G=M-TS&&=-V_{d-2}r_+^{d-3}\left[\frac{P r_+^2}{(d-1)(d-2)}
+\frac{w_2}{2}\right.\nonumber\\
&&\left.+\frac{(d-3) w_3}{r_+}+\frac{3(d-3)(d-4) w_4}{2r_+^2}\right].\label{eq:20a}
\end{eqnarray}
Here $r_+$ is understood as a function of pressure and temperature,
$r_+=r_+(P,\hat{T})$, via equation of state (\ref{eos}).

\begin{figure}[htb]
\centering
\subfigure[]{\label{fig:a} 
  \includegraphics{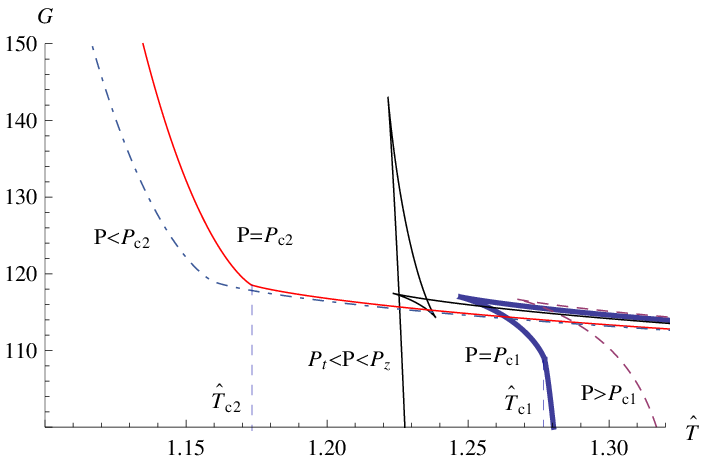}}%
\hfill%
\subfigure[]{\label{fig:b} 
  \includegraphics{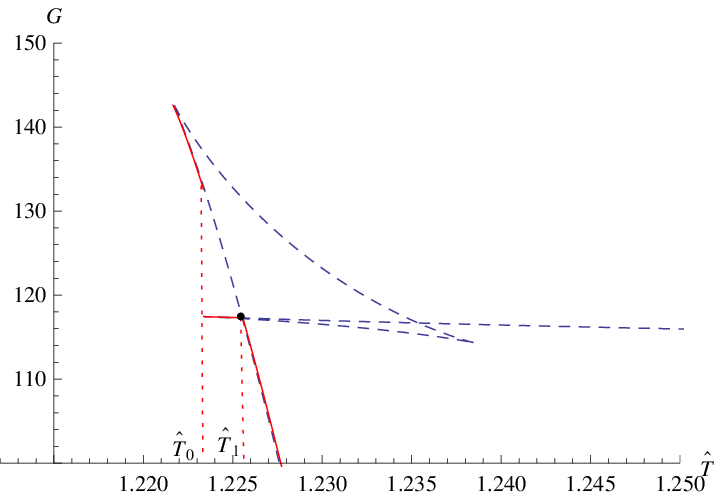}}%
\caption{ \textbf{The $G-\hat{T}$ diagrams of six-dimensional AdS black holes for $w_2=-1$,
$w_3=1$ and $w_4=-0.7$.}
For $P\in(P_t, P_z)$, we observe a ``zeroth-order phase transition''
signifying the onset of an RPT in Fig.{\ref{fig3}}(a). In Fig.{\ref{fig3}}(b)
with $P=0.0869\in(P_t, P_z)$,
a close-up of Fig.{\ref{fig3}}(a) illustrates the discontinuity in the global
minimum of $G$ at $\hat{T}=\hat{T}_0\approx1.223398\in(\hat{T}_t, \hat{T}_z)$
and the so-called Van der Waals-like phase transition at
$\hat{T}=\hat{T}_1\approx1.22554$.}\label{fig3}
\end{figure}

In the range of $\frac{3(d-4)w_3^2}{8(d-5)w_4}<w_2<\frac{3(d-4)w_3^2}{9(d-5)w_4}$
with $w_3>0$ and $w_4<0$, the behavior of $G$ in the six-dimensional spacetime
is depicted in Fig.~{\ref{fig3}}(a).
We have one physical (with positive pressure) critical point and the corresponding
first order SBH/LBH phase transition. This phase transition occurs
for $\hat{T}<\hat{T}_{c1}$ and terminates at $\hat{T}=\hat{T}_t$.
In particular, there also exists a certain range of temperatures,
$\hat{T}\in(\hat{T}_t, \hat{T}_z)$, for which the global minimum of
$G$ is discontinuous; see Fig.{\ref{fig3}}(b).
In this range of temperatures, two separate branches
of intermediate size and small size black holes co-exist. They are separated by a
finite jump in $G$, which is so-called ``zeroth-order phase transition''.
This phenomenon is also seen in superfluidity and superconductivity \cite{Maslov}.

\begin{figure}[htb]
\centering
\subfigure[]{\label{fig:a} 
  \includegraphics{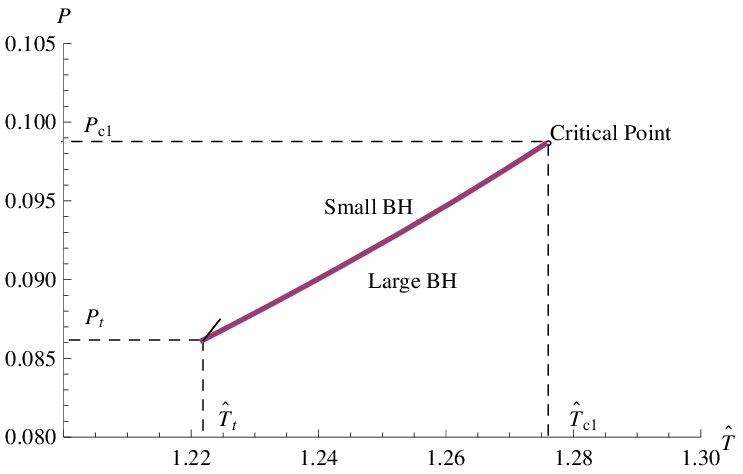}}%
\hfill%
\subfigure[]{\label{fig:b} 
  \includegraphics{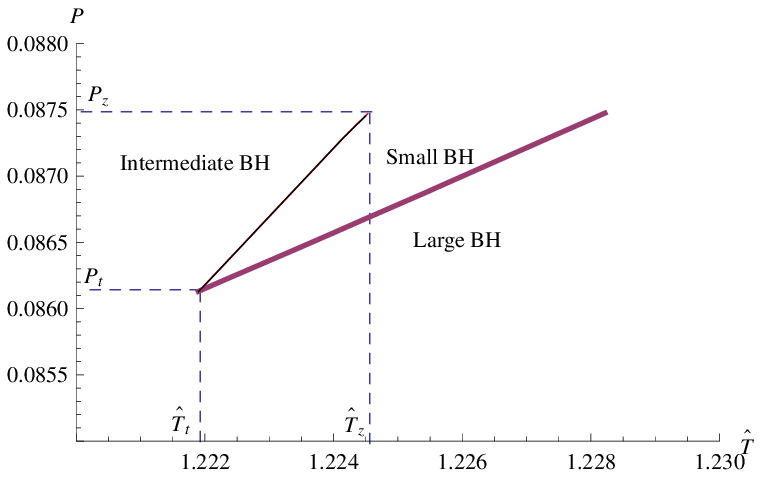}}%
  \caption{\textbf{The $P-\hat{T}$ diagram of six-dimensional AdS black holes for
  $w_2=-1$, $w_3=1$ and $w_4=-0.7$}.
The co-existence line of Van der Waals-like phase transition
is depicted by a thick solid line, which initiates
from the critical point $(P_{c1}, \hat{T}_{c1})$ and terminates at $(P_t, \hat{T}_t)$.
The solid line in the inset indicates the co-existence line
of small and intermediate black holes, separated by a finite
gap in $G$, indicating the RPT.
It commences from $(P_z, \hat{T}_z)$ and terminates at $(P_t, \hat{T}_t)$.}\label{fig4}
\end{figure}

This novel situation can also clearly be illustrated in the $P-\hat{T}$
diagrams in Fig.{\ref{fig4}}. There is the expected SBH/LBH line
of co-existence, which initiates from the critical point $(\hat{T}_{c1}, P_{c1})$
and terminates at $(\hat{T}_t, P_t)$. Especially, a ``triple point''
between the small, intermediate, and large
black holes appears in the point $(\hat{T}_t, P_t)$.
For $\hat{T}\in(\hat{T}_t, \hat{T}_z)$, a new IBH/SBH line of
coexistence appears and then it terminates in another critical point
$(\hat{T}_z, P_z)$. The range for the RPT is quite narrow and must
be determined numerically. Taking $w_2=-1$, $w_3=1$, $w_4=-0.7$ and
$d=6$, we obtain
\begin{eqnarray}
&&(\hat{T}_t, \hat{T}_z, \hat{T}_{c1})\approx(1.22194, 1.22459,1.27718),\nonumber\\
&&(P_t, P_z, P_{c1})\approx(0.08615, 0.08747,0.09899).
\end{eqnarray}

\begin{figure}[htb]
\centering
\subfigure[$w_2=-1$, $w_3=1$ and $w_4=0.1$]{\label{fig:a} 
  \includegraphics[width=0.3\textwidth]{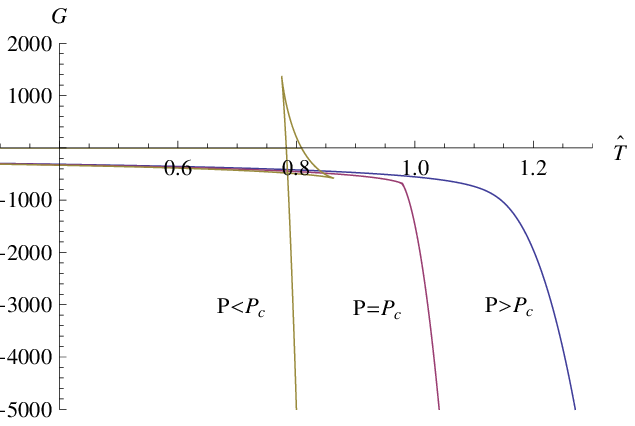}}%
\hfill%
\subfigure[$w_2=-1$, $w_3=-1$ and $w_4=0.6$]{\label{fig:b} 
  \includegraphics[width=0.3\textwidth]{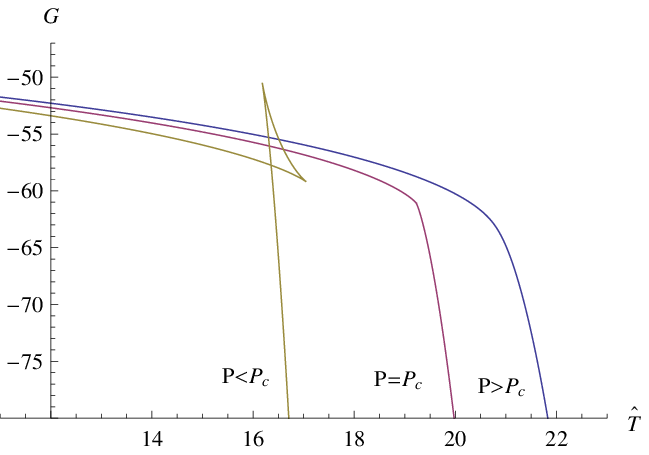}}%
  \hfill%
\subfigure[$w_2=1$, $w_3=-1$ and $w_4=0.5$]{\label{fig:c} 
  \includegraphics[width=0.3\textwidth]{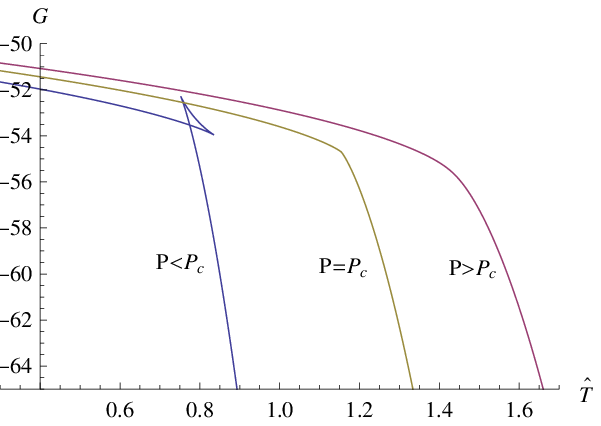}}%
\caption{ \textbf{The $G-\hat{T}$ diagrams of six-dimensional AdS black holes.}
The behavior of the Gibbs free energy is depicted as a function of temperature
for fixed pressure. There is one critical point
and the corresponding Van der Waals-like phase transition
for $\hat{T}<\hat{T}_c$. }\label{fig5}
\end{figure}

In Fig.~{\ref{fig5}}, we also plot the behavior of Gibbs free energy of six-dimensional
AdS black hole for three other cases, showed in Table~\ref{sphericcase}.
One can see that the $G$ surface demonstrates the characteristic
``swallow tail'' behavior, which indicates the occurrence of Van der Waals-like
SBH/LBH phase transition when $P<P_c$ in the corresponding system.
Moreover, the corresponding $P-\hat{T}$ diagram (not shown) is reminiscent of what was
observed for charged black holes in \cite{Kubiznak:2012wp} and is analogous
to the Van der Waals $P-\hat{T}$ diagram in each case.

\begin{figure}[htb]
\centering
\subfigure[4d SAdS black hole]{\label{fig:a} 
  \includegraphics{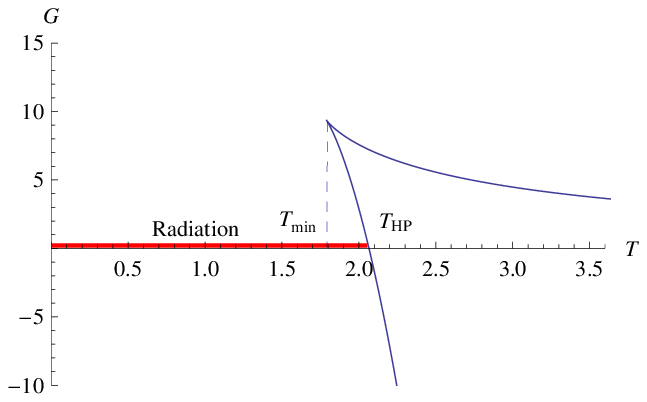}}%
\hfill%
\subfigure[5d SAdS black hole]{\label{fig:b} 
  \includegraphics{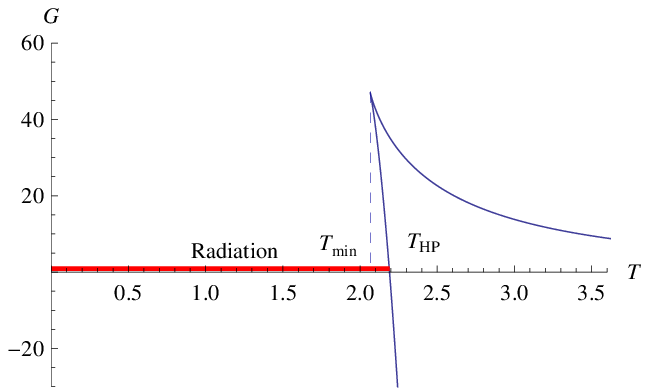}}%
\caption{ \textbf{The $G-T$ diagrams of four and five dimensional Schwarzschild AdS
black holes for $P=0.2$ and $k=1$.}
The radiation phase is displayed by horizontal magenta line.
The Hawking-Page phase transition between thermal radiation and
black holes occurs at $T=T_{HP}$. For $T>T_{HP}$, this branch
has negative Gibbs free energy and the corresponding black holes
represent the globally thermodynamically preferred state.}\label{fig6}
\end{figure}

When $m\rightarrow0$, namely, $w_3=w_4=0$ and $w_2=-\frac{k}{8\pi}$,
we find that the free energy $G$ always maintains negative in cases of $k=0$ and $k=-1$,
which correspond to a Ricci flat and hyperbolic topology of the
black hole horizon of $d$-dimensional Schwarzschild AdS black hole, respectively.
It is of great interest to discuss d-dimensional Schwarzschild AdS black hole
with spherical horizon ($k=1$). In Fig.~{\ref{fig6}}, it is shown that the
temperature $T$ has a minimal value $T_{min}$ below which no black hole solution exists.
When the temperature drops to a certain value larger than $T_{min}$,
the Gibbs free energy $G$ will be larger than zero, and then a more
stable vacuum will take place. At $T=T_{HP}$, there is a first order
Hawking-Page \cite{Hawking:1982dh} phase transition between thermal radiation and
black hole phase. This phase transition can be interpreted as a
confinement/deconfinement phase transition in the dual quark
gluon plasma \cite{Witten:1998zw}.

\section{Closing remarks}
\label{4s}

In this paper we have studied the thermodynamic behavior of higher-dimensional
AdS black hole in the dRGT massive gravity. We discussed this
issue in the extended phase space where the cosmological
constant appears as the pressure of the thermodynamic system and its conjugate
quantity is the thermodynamic volume of the black holes.
Following the standard thermodynamic techniques, we have written out the
equations of state and examined the phase structures.
When the coupling coefficients of massive potential satisfy $\frac{3(d-4)w_3^2}{8(d-5)w_4}<w_2<\frac{3(d-4)w_3^2}{9(d-5)w_4}$
with $w_3>0$ and $w_4<0$, we found that a monotonic lowering of
the temperature yields a large-small-large black hole transition,
where we refer to the latter ``large'' state as an intermediate
black hole (IBH), which is reminiscent of reentrant phase transitions.
Moreover, this process is also accompanied by a discontinuity
in the global minimum of the Gibbs free energy, referred
to as a zeroth-order phase transition.
For three other cases in Table~\ref{sphericcase},
the usual Van der Waals-like small/large black hole phase transition
occurred when coupling coefficients of massive potential adopt
some proper values in the higher-dimensional spacetime.
In addition, the solution (\ref{solution}) recovered d-dimensional
Schwarzschild AdS black holes in case of $m\rightarrow0$. It demonstrated the
existence of so-called Hawking-Page phase transition between
SAdS black hole with spherical horizon and vacuum if $d\geq4$.

It is necessary to point out that the charged black hole \cite{Hendi:2015pda},
Born-Infeld black hole \cite{Hendi:2016yof}, and black hole in the
Maxwell and Yang-Mills fields \cite{Meng:2016its}
have been recently constructed in Gauss-Bonnet massive gravity.
It has also showed the existence of
Van der Waals like first order SBH/LBH phase transition in these models.
It would be interesting to extend our discussion to these
black hole solutions and find out whether the reentrant
phase transition can appear.

\begin{acknowledgements}
The work is supported by the National Natural Science Foundation
of China (Grant No.11605152, No.11275099, and No.11647050),
and Natural Science Foundation of Jiangsu Province under Grant No.BK20160452.
\end{acknowledgements}

\end{document}